\documentclass[aps,prd,preprint]{revtex4-1}
\usepackage {graphicx}
\usepackage[toc,page]{appendix}
\usepackage{amsmath}
\usepackage{tikz}
\newcommand{\be}{\begin{equation}}
\newcommand{\ee}{\end{equation}}
\begin{document}
\title{Gravitational wave memory and the wave equation}
\author{David Garfinkle}
\email{garfinkl@oakland.edu}
\affiliation{Dept. of Physics, Oakland University, Rochester, MI 48309, USA}
\affiliation{Leinweber Center for Theoretical Physics, Randall Laboratory of Physics, University of Michigan, Ann Arbor, MI 48109-1120, USA}

\date{\today}

\begin{abstract}
Gravitational wave memory and its electromagnetic analog are shown to be straightforward consequences of the wave equation.  From Maxwell's equations one can derive a wave equation for the electric field, while from the Bianchi identity one can derive a wave equation for the Riemann tensor in linearized gravity.  Memory in both cases is derived from the structure of the source of those wave equations.

\end{abstract}


\maketitle

\section{Introduction}

Gravitational wave memory is a residual effect on the gravitational wave detector left after the gravitational wave has passed.  It was first found in linearized gravity by Zeldovich and Polnarev\cite{zeldovich} and then in full nonlinear general relativity by Christodoulou.\cite{christodoulou} That gravitational radiation has such residual effects is already surprising; but perhaps even more surprising is that the amount of the memory is encoded in the asymptotic behavior of the escaping matter\cite{zeldovich} and energy.\cite{christodoulou}  

In \cite{thorne} gravitational wave memory was calculated and expressed in terms of the transverse-traceless part of the metric perturbation.  In \cite{strominger} a relation between memory and BMS supertranslations is calculated.  While the work in \cite{thorne} and \cite{strominger} is entirely reasonable, it has given rise to somewhat odd and narrow views.  Under the influence of \cite{thorne} a widespread view has developed that gravitational wave memory is nothing but a property of the transverse traceless metric perturbation and can only be understood in this way.  And under the influence of \cite{strominger} an even more widespread view has developed that gravitational wave memory is nothing but a property of BMS symmetry, and can only be understood in this way.

As an alternative to the points of view generated by the work of  \cite{thorne} and \cite{strominger}, we note that the main properties of memory can be explained in terms of simple properties of the flat spacetime wave equation.  The relevant properties of the wave equation are given in section \ref{waveEqn}.  Then these properties are used to treat the electromagnetic analog of gravitational wave memory\cite{EMmemory} in section \ref{EM}.  Gravitational wave memory is treated in section \ref{GW} and conclusions are given in section \ref{Conclude}.

\section{Wave Equation}
\label{waveEqn}

Let $\Psi$ satisfy the Minkowski spacetime wave equation
\begin{equation}
{\partial ^a}{\partial _a} \Psi = - 4 \pi S
\label{wave}
\end{equation} 
for some source term $S$.  The retarded solution of eqn. (\ref{wave}) is
\begin{equation}
\Psi(t,{\vec x}) = \int \; {d^3} y \; {\frac {S({t_r} ,{\vec y} )} {| {\vec x} - {\vec y}|}} \; \; \; .
\label{ret}
\end{equation} 
Here the retarded time $t_r$ is given by the relation
\begin{equation}
t - {t_r} = | {\vec x} - {\vec y}|   \; \; \; .
\label{tret}
\end{equation}
That is, the integration in eqn. (\ref{ret}) is done over the past light cone of the point $(t,{\vec x})$.  

Now we would like to know the behavior of $\Psi$ at large distances in outgoing null directions.  We therefore suppose that ${\vec x} = r {\hat r}$ where $r$ is large and $\hat r$ is a unit vector.  We also introduce the null coordinate $u$ given by $u = t-r$.  As long as at all places where the source is nonzero we have $|{\vec y}| \ll r$, we can approximate $1/|{\vec x} - {\vec y} | $ by $1/r$.  Thus eqn. (\ref{ret}) becomes
\begin{equation}
\Psi(u,r,{\hat r}) = {\frac 1 r} \; \int \; {d^3} y \; S({t_r},{\vec y}) \; \; \; ,
\label{ret2}
\end{equation}
while eqn. (\ref{tret}) becomes 
\begin{equation}
{t_r} = u + r - |r{\hat r} - {\vec y}|
\label{tret2}
\end{equation}

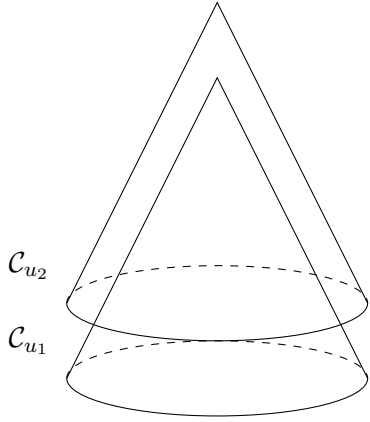
\begin{figure}
\centering
\begin{tikzpicture}
\draw (-2,0) arc (180:360:2cm and 0.5cm) -- (0,4) -- cycle;
\node at (-2.5,0.5) {${\cal C}_{u_2}$};
\draw[dashed] (-2,0) arc (180:0:2cm and 0.5cm);
\draw (-2,-1) arc (180:360:2cm and 0.5cm) -- (0,3) -- cycle;
\node at (-2.5,-0.5) {${\cal C}_{u_1}$};
\draw[dashed] (-2,-1) arc (180:0:2cm and 0.5cm);
\end{tikzpicture}
\caption{Past lightcones ${\cal C}_{u_2}$ with vertex $({u_2},r,{\hat r})$ (upper cone) and 
${\cal C}_{u_1}$ with vertex $ ({u_1},r,{\hat r}) $(lower cone).  The spacetime volume $\cal M$ being integrated over is the region between the two cones. (Note: despite the limitations of the figure, each cone extends infinitely to the past)}
\label{cones}
\end{figure}

Now for any interval $({u_1},{u_2})$, integrate eqn. (\ref{ret2}) over this interval to obtain
\begin{equation}
{\int _{u_1} ^{u_2}} \; du \; \Psi(u,r,{\hat r}) = {\frac 1 r} \; {\int _{u_1} ^{u_2}} \; du \; \int \; {d^3} y \; S({t_r},{\vec y}) \; \; \; .
\label{mem1}
\end{equation}
However, the Jacobian of the transformation between the $(u,{\vec y})$ coordinates and the $({t_r},{\vec y})$ coordinates is unity, so we can also write eqn. (\ref{mem1}) as 
\begin{equation}
{\int _{u_1} ^{u_2}} \; du \; \Psi(u,r,{\hat r}) = {\frac 1 r}  {\int _{\cal M}}\; {d^4} y \; S({t_r},{\vec y}) \; \; \; .
\label{mem2}
\end{equation}
Here the integral is over the spacetime region $\cal M$ between the past light cones of the points $({u_2},r,{\hat r})$ and $({u_1},r,{\hat r})$ (see figure \ref{cones}).  

Now, we specialize to the case where $S$ is a total divergence, that is where there is a current $Q^a$ for which $S = {\partial _a}{Q^a}$.  Then denoting the flat spacetime volume element by $\epsilon_{abcd}$ we have 
\begin{equation}
{\int _{\cal M}} \; S {\epsilon_{abcd}} = {\int _{\cal M}} \; {\partial _e}{Q^e} {\epsilon_{abcd}} = 
{\int_{\partial {\cal M}}} \; {Q^a}{n_a} {{\tilde \epsilon}_{bcd}}
\label{Gauss}
\end{equation}
Here $\partial {\cal M}$ is the boundary of $\cal M$, $n_a$ is a normal to $\partial {\cal M}$, and 
${\tilde \epsilon}_{bcd}$ is a volume element on $\partial {\cal M}$ satisfying
\begin{equation}
{\epsilon_{abcd}} = 4 {n_{[a}}{{\tilde \epsilon}_{bcd]}}
\label{nullGauss}
\end{equation}
The reason for this somewhat complicated statement of Gauss' theorem is that the boundary is a null surface (see {\it e.g.} Appendix B of \cite{wald}).  Using eqn. (\ref{Gauss}) in eqn. (\ref{mem2}) we obtain
\begin{equation}
{\int _{u_1} ^{u_2}} \; du \; \Psi(u,r,{\hat r}) = {\frac 1 r} \left [ {\int _{C_{u_2}}} \;  {Q^a}{n_a} {{\tilde \epsilon}_{bcd}} \; - \; {\int _{C_{u_1}}} \;  {Q^a}{n_a} {{\tilde \epsilon}_{bcd}} \right ] \; \; \; .
\label{mem3}
\end{equation}
Here $C_{u_1}$ (resp. $C_{u_2}$) is the past light cone of the point $({u_1},r, {\hat r})$ 
(resp. $({u_2},r, {\hat r})$), and we have chosen the past pointing null normal vector for both the integral over $C_{u_1}$ and the integral over $C_{u_2}$.

Note that given our assumption that $|{\vec y}| \ll r$, it follows that $t_r$ in eqn. (\ref{tret2}) is well approximated by 
\begin{equation}
{t_r} = u + {\hat r} \cdot {\vec y} \; \; \; .
\label{tret3}
\end{equation}
That is, our integrals over null cones become integrals over null planes. (This is just the well known property of Minkowski spacetime that the limit as the vertex tends to null infinity of a null cone is a null plane).  Thus eqn. (\ref{mem3}) becomes
\begin{equation}
{\int _{u_1} ^{u_2}} \; du \; \Psi(u,r,{\hat r}) = {\frac 1 r} \left [ {\int _{P_{u_2}}} \;  {Q^a}{n_a} {{\tilde \epsilon}_{bcd}} \; - \; {\int _{P_{u_1}}} \;  {Q^a}{n_a} {{\tilde \epsilon}_{bcd}} \right ] \; \; \; .
\label{mem4}
\end{equation}
Here $P_{u_1}$ is the null plane given by ${t_r} = {u_1} + {\hat r} \cdot {\vec y}$ and correspondingly for $P_{u_2}$.  

Note that from eqn. (\ref{tret3}) it follows that the components of the normal vector are given by
\begin{equation}
{n_0}=1 \; , \; \; \; {n_i}= - {{\hat r}_i} \; \; \; ,
\label{normal}
\end{equation}
and therefore from eqn. (\ref{nullGauss}) that
\begin{equation}
{{\tilde \epsilon}_{123}}=1
\end{equation}

We now check the consistency of our assumption that $|{\vec y}| \ll r$.   Suppose that the source $S$ is concentrated along the world line of an object or light ray with velocity $\vec v$.  That is a world line satisfying ${\vec y} = {{\vec y}_0} + {\vec v}{t_r}$.  Then we find from eqn. (\ref{tret3}) that 
\begin{equation}
{\vec y} = {{\vec y}_0} \; + \; {\frac {{\vec v} (u + {\hat r}\cdot {{\vec y}_0})} {1 - {\hat r}\cdot {\vec v}}}
\label{consistency}
\end{equation}  
Thus we find that our assumption is consistent provided that the quantity $1 - {\hat r}\cdot {\vec v}$ does not vanish.  For objects traveling slower than light, $1 - {\hat r}\cdot {\vec v} > 0$ so our assumption is consistent.  Even for objects traveling at the speed of light, $1 - {\hat r}\cdot {\vec v} > 0$ provided that $\vec v$ does not point in the direction of $\hat r$.  Thus in order to use the results of this section, we must check that any null part of the source $S$ vanishes whenever it points in the $\hat r$ direction.

\section{Electromagnetic Memory}
\label{EM}

The electromagnetic analog of gravitational wave memory\cite{EMmemory} is a kick ({\it i.e.} change in velocity) received by a test charge.  Since a test charge of charge $q$ and mass $m$ satisfies the equation of motion
\begin{equation}
m {\ddot {\vec x}} = q {\vec E} \; \; \; ,
\end{equation}
It follows that the kick received by the test charge is
\begin{equation}
\Delta {\vec v} = {\frac q m} {\int _{-\infty} ^\infty} \; {\vec E} \; dt \; \; \; .
\end{equation}

To apply the method of section \ref{waveEqn} we need to find an equation for the components of the electric field that is a wave equation with source.  Recall that Maxwell's equations are 
\begin{eqnarray}
{\partial _\alpha}{F_{\beta \gamma }} + {\partial _\beta}{F_{\gamma \alpha }} + {\partial _\gamma}{F_{\alpha \beta }} = 0 \; \; \; ,
\label{Maxwell1}
\\
{\partial ^\alpha}{F_{\alpha \beta}} = - 4 \pi {j_\beta} \; \; \; .
\label{Maxwell2}
\end{eqnarray}
Here $F_{\alpha \beta}$ is the Faraday tensor and $j_\alpha$ is the electromagnetic current four-vector.
Applying $\partial ^\gamma$ to eqn. (\ref{Maxwell1}) and using eqn. (\ref{Maxwell2}) we obtain
\begin{equation}
{\partial ^\gamma}{\partial _\gamma} {F_{\alpha \beta}} = - 4 \pi ({\partial _\alpha}{j_\beta} - {\partial _\beta}{j_\alpha} )
\label{waveF}
\end{equation}
Thus, each Cartesian component of the Faraday tensor is a solution of the wave equation.  In particular, the $i$ component of the electric field is 
$ {E_i} = {F_{i0}}$ so we find
\begin{equation}
{\partial ^\gamma}{\partial _\gamma} {E_i} = - 4 \pi ({\partial _i}{j_0} - {\partial _0}{j_i} )
\label{waveE}
\end{equation}
The right hand side of eqn. (\ref{waveE}) is of the form $-4\pi {\partial _a}{Q^a}$ where the current $Q^a$ is given by
\begin{equation}
{Q^\mu} = {{\delta ^\mu} _i} {j_0} - {{\delta ^\mu} _0} {j_i} \; \; \; .
\label{EMmemoryCurrent}
\end{equation}
Using eqn. (\ref{normal}) we have 
\begin{equation}
{Q^a}{n_a} = - ({{\hat r}_i}{j_0} + {j_i})
\label{EMQ}
\end{equation}

Note that there is a different $Q^\mu$ for each component of the Faraday tensor.  That is there is a tensor ${\cal Q}^\mu _{\alpha \beta}$ such that 
\begin{equation}
{\partial ^\gamma}{\partial _\gamma} {F_{\alpha \beta}} = - 4 \pi {\partial _\mu} {{\cal Q}^\mu _{\alpha \beta}} \; \; \; .
\label{tensorQ}
\end{equation}
Here the tensor ${\cal Q}^\mu _{\alpha \beta}$ is given by 
\begin{equation}
{{\cal Q}^\mu _{\alpha \beta}} = {{\delta ^\mu}_\alpha} {j_\beta} - {{\delta ^\mu}_\beta} {j_\alpha} \; \; \; .
\end{equation}
The $Q^\mu$ in eqn. (\ref{EMmemoryCurrent}) is ${{\cal Q}^\mu _{i0}} $.  Nonetheless, for simplicity of expression we will stick with the notation of eqn. (\ref{EMmemoryCurrent}) rather than writing our expressions in terms of ${\cal Q}^\mu _{\alpha \beta}$.

Define the projection operator $P_{ij}$ by
\begin{equation}
{P_{ij}} \equiv {\delta_{ij}} - {{\hat r}_i}{{\hat r}_j}
\label{ProjectDef}
\end{equation}
That is, $P_{ij}$ projects into the space orthogonal to $\hat r$.  
Then we find from eqn. (\ref{EMQ}) that
\begin{equation}
{Q^a}{n_a} =  {{\hat r}_i}(-{j_0} - {\hat r}\cdot {\vec j}) - {{P_i}^k} {j_k} = 
{{\hat r}_i}({j^\mu}{n_\mu}) - {{P_i}^k} {j_k}
\label{EMQ2}
\end{equation}

However $\int {j^\mu}{n_\mu}$ is the total charge and is thus the same quantity whether the integral is done over the null plane $P_\infty$ or the null plane $P_{-\infty}$.  This can be seen more clearly in terms of the null cones ${\cal C}_{u_1}$ and ${\cal C}_{u_2}$ of figure (\ref{cones}).  Since ${\partial _\mu}{j^\mu} =0$ it follows that $\int _{\cal M} {\partial _\mu}{j^\mu} =0$ where 
$\cal M$ is the spacetime region bounded by the null cones ${\cal C}_{u_1}$ and ${\cal C}_{u_2}$.  Then using Gauss' theorem (and since we are assuming that no charge is coming in from past null infinity) it follows that $\int_{{\cal C}_{u_2}} {j^\mu}{n_\mu} =  \int_{{\cal C}_{u_1}} {j^\mu}{n_\mu} $.  However the null plane $P_\infty$ is just the limit of the null cone ${\cal C}_{u_2}$ as first $r \to \infty $ and then ${u_2} \to \infty$.  Similarly, the null plane $P_{-\infty}$ is just the limit of the null cone ${\cal C}_{u_1}$ as first $r \to \infty $ and then ${u_1} \to -\infty$. It then follows that $\int_{P_\infty} {j^\mu}{n_\mu} =  \int_{P_{-\infty}} {j^\mu}{n_\mu} $.  Note that each of the quantities $\int_{P_\infty} {j^\mu}{n_\mu}$ and $\int_{P_{-\infty}} {j^\mu}{n_\mu}$ is nonzero.  However, they are the {\emph {same}} nonzero quantity.  And since in eqn. (\ref{mem4}) these quantities contribute to the memory with opposite sign, it then follows that the total contribution to the memory from both quantities together is zero.

Thus applying eqn. (\ref{EMQ2}) to eqn. (\ref{mem4})
we find that the electromagnetic memory is given by 
\begin{equation}
{\int _{-\infty} ^\infty} \; {E_i} \; dt = {\frac 1 r} \; 
\left [ - {\int _{P_\infty}} \; {{P_i}^k} {j_k} \; {d^3}y \; + \; {\int _{P_{-\infty}}} \; {{P_i}^k} {j_k} \; {d^3}y \right ] \; \; \; .
\label{EMmem1}
\end{equation}

As pointed out in \cite{Istvan,Abhay1,Abhay2} There are two different definitions of transverse fields used to describe radiation: a local algebraic definition involving projection using $P_{ij}$ and a nonlocal definition using divergence-free tensors.  In our treatment, it is the simple, local, algebraic definition that is relevant for memory. 

As in \cite{EMmemory} we will assume the following behavior of the charges at early and at late times: at early times the sources of charge consist of isolated bodies traveling at constant velocity.  At late times the sources of charge consist both of such bodies and of an outgoing radiation of charge that travels along null directions (as happens for the massless Maxwell-Klein-Gordon equation\cite{Bieri}).  Now consider a single isolated body with charge $e$ and four-velocity ${u^\mu}=\gamma (1,{\vec v})$.  The integral over the null plane of $j^\mu$ must point in the direction of $u^\mu$ and yield $e$ when contracted with $n_\mu$.  It then follows that
\begin{equation}
{\int _{{\cal P}_u}}\; {d^3}y \; {j^\mu} = {\frac {e {u^\mu}} {\gamma ( 1 - {\hat r}\cdot {\vec v})}} \; \; \; .
\label{EMbody1}
\end{equation}
We then find that 
\begin{equation}
 {\int _{{\cal P}_u}}\; {d^3}y \; {{P_i}^k} {j_k} = {\frac {e {{P_i}^k} {v_k}} { 1 - {\hat r}\cdot {\vec v}}} \; \; \; .
\label{EMbody2}
\end{equation}

We now consider the contribution of the outgoing radiation of charge to the memory.  As before, we use the notation $t_r$ for the time coordinate of the $y$ coordinate system.  However, we will use the notation ${r_y}, {{\hat r}_y}$ and $d {\Omega_y}$ to denote respectively the radial coordinate, radial unit vector, and solid angle of the $y$ coordinate system.  Then at late times the current of the outgoing radiation of charge takes the form
\begin{equation}
{j^\mu} = {r_y ^{-2}} {L_q} ({t_r}-{r_y},{{\hat r}_y}) {\ell ^\mu} \; \; \; .
\label{jNull}
\end{equation}
Here the null vector $\ell ^\mu$ is ${\ell ^\mu}={t^\mu}+{{\hat r}_y ^\mu}$ where $t^\mu$ is the unit vector in the time direction.  The quantity $L_q$ is the charge radiated per unit solid angle per unit time.  Define the quantity $F_q$ by
\begin{equation}
{F_q}({{\hat r}_y}) \equiv {\int _{-\infty} ^\infty} \;  {L_q} (s,{{\hat r}_y}) \; ds \; \; \; .
\label{FqDef}
\end{equation}
Then ${F_q}({{\hat r}_y})$ is the total charge radiated per unit solid angle.  Now consider the null plane ${{\cal P}_{u_2}}$ with $u_2$ large and positive.  On that null plane we have ${t_r} = {u_2} + {r_y}{\hat r}\cdot {{\hat r}_y}$ so applying eqn. (\ref{jNull}) we obtain
\begin{equation}
 {\int _{{\cal P}_{u_2}}}\; {d^3}y \; {{P_i}^k} {j_k} = \int \; d {\Omega _y} \; {\int _0 ^\infty} \; {r_y ^2} \; d {r_y} \;  {r_y ^{-2}} {L_q} ({u_2}-{r_y}(1-{\hat r}\cdot {{\hat r}_y}),{{\hat r}_y})
 {P_{ik}}{{\hat r}_y ^k} \; \; \; .
 \label{NullMem1}
 \end{equation}
 Then changing the variable of integration from $r_y$ to $s \equiv {u_2}-{r_y}(1-{\hat r}\cdot {{\hat r}_y})$ and applying eqn. (\ref{FqDef}) we obtain
\begin{equation}
 {\int _{{\cal P}_{u_2}}}\; {d^3}y \; {{P_i}^k} {j_k} = \int \; d {\Omega _y} \; {\frac {{F_q}({{\hat r}_y}) \, {P_{ik}}{{\hat r}_y ^k}} {1 - {\hat r}\cdot {{\hat r}_y}}}
\label{NullMem2}
\end{equation} 
Note that despite the fact that the denomenator can vanish, the integrand in eqn. (\ref{NullMem2}) is non-singular, because the denominator only vanishes when ${{\hat r}_y}={\hat r}$ and that is also where the numerator vanishes.  Also note that for the case where ${F_q}({{\hat r}_y})$ is a constant, the integral in eqn. (\ref{NullMem2}) vanishes.  Thus if one considers an expansion of 
${F_q}({{\hat r}_y})$ in spherical harmonics, it is only the $\ell >0$ spherical harmonics that contribute to the memory.

Finally applying the results of eqn. (\ref{EMbody2}) and eqn. (\ref{NullMem2}) to eqn. (\ref{EMmem1}) we find that the total electromagnetic memory is 
\begin{equation}
{\int _{-\infty} ^\infty} \; {E_i} \; dt = {\frac 1 r} \; 
\left [ - \; \int \; d {\Omega _y} \; {\frac {{F_q}({{\hat r}_y}) \, {P_{ik}}{{\hat r}_y ^k}} {1 - {\hat r}\cdot {{\hat r}_y}}} \; -\; {\sum _{\rm late}} {\frac {{e_{(\ell)}} {{P_i}^j} {v_{(\ell)j}}} { 1 - {\hat r}\cdot {\vec v}_{(\ell)}}} \; + \; {\sum _{\rm early}} {\frac {{e_{(\ell)}} {{P_i}^j} {v_{(\ell)j}}} { 1 - {\hat r}\cdot {\vec v}_{(\ell)}}} \right ] \; \; \; .
\label{EMmem2}
\end{equation}
Here the first sum is over all late time isolated bodies with charges $e_{(\ell)}$ and velocities ${\vec v}_{(\ell)}$, while the second sum is the corresponding sum for the early time isolated bodies.

\section{Gravitational Wave Memory}
\label{GW}
For two nearby objects in free fall, their separation $s^i$ satisfies
\begin{equation}
{{\ddot s}_i} =  {R_{i00j}}{s^j} \; \; \; ,
\label{Jacobi}
\end{equation}
where the $0$ direction is the rest frame of the objects.  We apply eqn. (\ref{Jacobi}) in the asymptotic region, {\it i.e.} at large distances in null directions.  Then the permanent change in separation
$\Delta {s^i}$ satisfies
\begin{equation}
{\Delta {s_i}} = {M_{ij}}(r,{\hat r}){s^j} \; \; \; ,
\end{equation}
where the tensor ${M_{ij}}(r,{\hat r})$ (which we will call the memory tensor) is given by 
\begin{equation}
{M_{ij}} (r,{\hat r}) = {\int_{-\infty} ^\infty} {V_{ij}} (u,r,{\hat r})\; du \; \; \; ,
\label{MemoryTensor}
\end{equation}
with the tensor ${V_{ij}} (u,r,{\hat r})$ (which we will call the velocity tensor) given by
\begin{equation}
 {V_{ij}}(u,r,{\hat r}) = {\int_{-\infty} ^u} \; d {\tilde u} \; ( {R_{i00j}}({\tilde u,r,{\hat r}})) \; \; \; .
 \label{News}
\end{equation}

To apply the method of section \ref{waveEqn} we need to find an equation for the components of the Riemann tensor that is a wave equation with source.  We will treat only weak gravitational fields and work to linear order in perturbation theory.  Then the Bianchi identity is
\begin{equation}
{\partial _\alpha}{R_{\beta \gamma \delta \epsilon}} \; + \; 
{\partial _\beta}{R_{\gamma \alpha \delta \epsilon}} \; + \; 
{\partial _\gamma}{R_{\alpha \beta \delta \epsilon}} = 0 \; \; \; .
\label{Bianchi}
\end{equation}
Note that as in \cite{BG} using the perturbed Riemann tensor rather than the perturbed metric yields results that are manifestly gauge invariant, and where there is no need to choose a gauge.
Contracting eqn. (\ref{Bianchi}) with the Minkowski spacetime inverse metric $\eta ^{\alpha \epsilon}$ we obtain
\begin{equation}
{\partial ^\alpha}{R_{\beta \gamma \delta \alpha}} = {\partial _\gamma}{R_{\beta \delta}} \; - \; 
{\partial _\beta}{R_{\gamma \delta}} \; \; \; .
\label{Bianchi2}
\end{equation}
Then applying $\partial ^\alpha$ to eqn. (\ref{Bianchi}) and using eqn. (\ref{Bianchi2}) we obtain
\begin{equation}
{\partial ^\alpha}{\partial _\alpha}{R_{\beta \gamma \delta \epsilon}} = {\partial _\beta} \left ( {\partial _\delta}{R_{\epsilon \gamma}} \; - \; {\partial _\epsilon}{R_{\delta \gamma}} \right )
\; + \;  {\partial _\gamma} \left ( {\partial _\epsilon}{R_{\delta \beta}} \; - \; {\partial _\delta}{R_{\epsilon \beta}} \right ) \; \; \; .
\label{RiemannWave}
\end{equation}
Thus each Cartesian component of the Riemann tensor is a solution of the wave eqation with source.  In particular we have
\begin{equation}
{\partial ^\alpha}{\partial _\alpha}{R_{i00j}} =  {\partial _i} \left ( {\partial _0}{R_{0j}} - {\partial _j}{R_{00}} \right ) \; + \; {\partial _0} \left ( {\partial _j}{R_{0i}} - {\partial _0}{R_{ij}}  \right )  \; \; \; \; .
\label{RiemannWave2}
\end{equation}
Thus we have $ {\partial ^\alpha}{\partial _\alpha}{R_{i00j}} = - 4 \pi {\partial _\mu}{Q^\mu}$ where
\begin{equation}
{Q^\mu} = {\frac 1 {4 \pi}} \; \left [ {{\delta ^\mu} _0} \left ( {\partial _0}{R_{ij}} - {\partial _j}{R_{0i}} \right ) \; + \; {{\delta ^\mu} _i} \left ( {\partial _j}{R_{00}} - {\partial _0}{R_{0j}} \right ) \right ]
\label{RiemannSource}
\end{equation}

Analogously to the electromagnetic case, there is a $Q^\mu$ for each component of the Riemann tensor.  That is, there is a tensor  ${\cal Q}^\mu _{\alpha \beta \gamma \delta}$ such that 
\begin{equation}
{\partial ^\gamma}{\partial _\gamma} {R_{\alpha \beta \gamma \delta}} = - 4 \pi {\partial _\mu} {{\cal Q}^\mu _{\alpha \beta \gamma \delta}} \; \; \; .
\label{tensorQgrav}
\end{equation}
Thus, we could write our expressions in terms of ${\cal Q}^\mu _{\alpha \beta \gamma \delta}$ but we will not do so.

Applying eqns. (\ref{mem3}), (\ref{RiemannWave2}) and (\ref{RiemannSource}) to eqn. (\ref{News}) we obtain
\begin{equation}
 {V_{ij}}(u,r,{\hat r}) = {\frac 1 r}  {\int _{C_u}} \; {d^3}y \; {\frac 1 {4 \pi}} \; \left [ 
 {\partial _0}{R_{ij}} - {\partial _j}{R_{0i}} - {{\hat r}_i} \left ( {\partial _j}{R_{00}} - {\partial _0}{R_{0j}} \right ) \right ] \; \; \; .
\label{News2}
\end{equation}
Here we have used the fact (see Appendix \ref{nokick}) that the source falls off sufficiently fast as $u \to - \infty$ so there is no contribution from the light cone at $u_1$ where ${u_1} \to - \infty$.  Similarly, the source falls off sufficiently fast at large $u$ that ${\lim _{u \to \infty}} {V_{ij}}(u,r,{\hat r}) = 0$.  Since the relative velocity of the two nearby objects is proportional to $V_{ij}$ this means that gravitational waves in asymptotically flat spacetime  do not produce a kick.  
Now integrating eqn. (\ref{News2}) we obtain
\begin{equation}
 {\int _{u_1} ^{u_2}} \; du \; {V_{ij}}(u,r,{\hat r}) = {\frac 1 r}  {\int _{\cal M}} \; {d^4}y \; {\frac 1 {4 \pi}} \; \left [ 
 {\partial _0}{R_{ij}} - {\partial _j}{R_{0i}} + {{\hat r}_i} \left ( {\partial _0}{R_{0j}}  -{\partial _j}{R_{00}} \right ) \right ] \; \; \; ,
\label{mem5}
\end{equation}
where as before $\cal M$ is the spacetime region between the past light cone of $({u_2},r,{\hat r})$ and the past light cone of $({u_1},r,{\hat r})$.  However, the integrand in eqn. (\ref{mem5}) is a total divergence: we have
\begin{equation}
{\frac 1 {4 \pi}} \; \left [   {\partial _0}{R_{ij}} - {\partial _j}{R_{0i}} + {{\hat r}_i} \left ( {\partial _0}{R_{0j}}  -{\partial _j}{R_{00}} \right ) \right ] = {\partial _\mu}{{\tilde Q}^\mu} \; \; \; ,
\end{equation}
where the current ${\tilde Q}^\mu$ is given by
\begin{equation}
{{\tilde Q}^\mu} = {\frac 1 {4\pi}} \; \left [ {{\delta ^\mu}_0}{R_{ij}} - {{\delta ^\mu} _j} {R_{0i}}  + {{\hat r}_i} \left (  {{\delta ^\mu}_0}{R_{0j}} - {{\delta ^\mu}_j}{R_{00}} \right ) \right ] \; \; \; ,
\label{Qt}
\end{equation}
and therefore we find that
\begin{equation}
{{\tilde Q}^\mu}{n_\mu} = {\frac 1 {4\pi}} \; \left [ {R_{ij}} + {R_{0i}}{{\hat r}_j} + 
{R_{0j}}{{\hat r}_i} + {R_{00}}{{\hat r}_i}{{\hat r}_j} \right ]
\label{Qtdotn}
\end{equation}
However, from the Einstein field equations we have ${R_{\mu\nu}}/(4\pi) = 2 {T_{\mu\nu}} - T {g_{\mu\nu}}$, which with some straightforward but tedious algebra allows us to rewrite eqn. (\ref{Qtdotn}) as ${{\tilde Q}^\mu}{n_\mu} = {S_1} + {S_2}$ where $S_1$ and $S_2$ are given by
\begin{eqnarray}
{S_1} &=&  \left ( {T^{0\mu}}{n_\mu} + {{T_m}^\mu}{n_\mu}{{\hat r}^m} \right ) ({\delta _{ij}} + {{\hat r}_i}{{\hat r}_j} ) \; - \; 2 {{T_i}^\mu}{n_\mu} {{\hat r}_j} \; - \; 2 {{T_j}^\mu}{n_\mu} {{\hat r}_i} 
\label{S1}
\\
{S_2} &=& 2 \left (  {{P_i}^m}{{P_j}^n} \; - \; {\textstyle {\frac 1 2}} {P^{mn}}{P_{ij}} \right ) {T_{mn}} \; \; \; .
\label{S2}
\end{eqnarray}
Here the projection operator ${{P_i}^m}{{P_j}^n}  -  (1/2) {P^{mn}}{P_{ij}}$
projects symmetric tensors into the space of symmetric trace-free tensors orthogonal to $\hat r$. 
 
The reason for the somewhat complicated looking decomposition of ${{\tilde Q}^\mu}{n_\mu}$ is that one can show that $S_1$ makes zero contribution to the memory.  The argument is essentially the same as the one in the previous section that showed that terms proportional to ${j^\mu}{n_\mu}$ made no net contribution to the electromagnetic memory.  This is because $ \int {T^{0\mu}}{n_\mu}$ is the total energy and $\int {T^{k\mu}}{n_\mu}$ is the total spatial momentum.  All terms in eqn. (\ref{S1}) thus give the same contribution whether the integral is done over the null plane $P_\infty$ or the null plane $P_{-\infty}$.  Since in eqn. (\ref{mem4}) the terms in the integral over $P_{-\infty}$ contribute with the opposite sign to the terms in the integral over $P_\infty$, it follows that the net contribution to the memory of $S_1$ is zero.

Therefore the memory tensor is given by 
\begin{eqnarray}
{M_{ij}} (r,{\hat r}) &=&  {\frac 2 r} \; \biggl [  {\int _{P_\infty}} \; {d^3}y \left ( 
 {{P_i}^m}{{P_j}^n}  -  {\textstyle {\frac 1 2}} {P^{mn}}{P_{ij}}
 \right ) {T_{mn}}
 \nonumber
 \\
&-& \; {\int _{P_{- \infty}}} \; {d^3}y \left (  {{P_i}^m}{{P_j}^n}  -  {\textstyle {\frac 1 2}} {P^{mn}}{P_{ij}}
 \right ) {T_{mn}} \biggr ] \; \; \; .
 \label{Gmem1}
\end{eqnarray}

At early times the stress-energy consists of isolated bodies traveling along timelike geodesics.  At late times the stress-energy consists of such isolated bodies along with outgoing radiation.  Consider the contribution of one isolated body to the memory.  Let $m$ be the mass of the body and ${u^\mu} = \gamma (1,{\vec v})$ be its four-velocity.  Then the integal of $T^{\mu \nu}$ over the null plane must be proportional to ${u^\mu}{u^\nu}$.  Furthermore, since the total energy of the body is $\gamma m$, it follows that the integeral over the null plane of ${T^{0\nu}}{n_\nu}$ must equal $\gamma m$.  It then follows that 
\begin{equation}
{\int _{{\cal P}_u}}\; {d^3}y \; {T^{\mu \nu}} = {\frac {m {u^\mu}{u^\nu}} {\gamma ( 1 - {\hat r}\cdot {\vec v})}} \; \; \; .
\label{body1}
\end{equation}
We therefore have
\begin{equation}
{\int _{{\cal P}_u}}\; {d^3}y \;  ( {{P_i}^m}{{P_j}^n}  -  {\textstyle {\frac 1 2}} {P^{mn}}{P_{ij}}
 ) {T_{mn}} = {\frac {m \gamma  ( {{P_i}^m}{{P_j}^n}  -  {\textstyle {\frac 1 2}} {P^{mn}}{P_{ij}}
  ) {v_m}{v_n}} { 1 - {\hat r}\cdot {\vec v}}} \; \; \; .
\label{body2}
\end{equation}

We now consider the contribution of outgoing radiation to the memory.  As before, we use the notation of $t_r$ for the time coordinate of the $y$ coordinate system, and ${r_y}, {{\hat r}_y}$ and $d {\Omega_y}$ to denote respectively the radial coordinate, radial unit vector, and solid angle of the $y$ coordinate system.  Then at late times the stress-energy of the outgoing radiation takes the form
\begin{equation}
{T^{\mu\nu}} = {r_y ^{-2}} L ({t_r}-{r_y},{{\hat r}_y}) {\ell ^\mu} {\ell ^\nu}\; \; \; ,
\label{TNull}
\end{equation}
where the null vector $\ell ^\mu$ is ${\ell ^\mu}={t^\mu}+{{\hat r}_y ^\mu}$ and $t^\mu$ is the unit vector in the time direction.  The quantity $L$ is the energy radiated per unit solid angle per unit time.  Define the quantity $F$ by
\begin{equation}
F({{\hat r}_y}) \equiv {\int _{-\infty} ^\infty} \;  L (s,{{\hat r}_y}) \; ds \; \; \; .
\label{FDef}
\end{equation}
Then $F({{\hat r}_y})$ is the total energy radiated per unit solid angle.  Now consider the null plane ${{\cal P}_{u_2}}$ with $u_2$ large and positive.  On that null plane we have ${t_r} = {u_2} + {r_y}{\hat r}\cdot {{\hat r}_y}$ so applying eqn. (\ref{TNull}) we obtain
\begin{eqnarray}
 {\int _{{\cal P}_{u_2}}}\; {d^3}y \;  ( {{P_i}^m}{{P_j}^n}  -  {\textstyle {\frac 1 2}} {P^{mn}}{P_{ij}}) {T_{mn}}  = 
 \nonumber
 \\
 \int \; d {\Omega _y} \; {\int _0 ^\infty} \; {r_y ^2} \; d {r_y} \;  {r_y ^{-2}} L ({u_2}-{r_y}(1-{\hat r}\cdot {{\hat r}_y}),{{\hat r}_y}) \; ( {P_{im}}{P_{jn}}  -  {\textstyle {\frac 1 2}} {P_{mn}}{P_{ij}}) {{\hat r}_y ^m}{{\hat r}_y ^n} 
  \; \; \; .
 \label{NullGMem1}
 \end{eqnarray}
 Then changing the variable of integration from $r_y$ to $s \equiv {u_2}-{r_y}(1-{\hat r}\cdot {{\hat r}_y})$ and applying eqn. (\ref{FDef}) we obtain
\begin{equation}
 {\int _{{\cal P}_{u_2}}}\; {d^3}y \;( {{P_i}^m}{{P_j}^n}  -  {\textstyle {\frac 1 2}} {P^{mn}}{P_{ij}}) {T_{mn}}  = \int \; d {\Omega _y} \; {\frac {F({{\hat r}_y}) \, ( {P_{im}}{P_{jn}}  -  {\textstyle {\frac 1 2}} {P_{mn}}{P_{ij}}) {{\hat r}_y ^m}{{\hat r}_y ^n}     } {1 - {\hat r}\cdot {{\hat r}_y}}}
\label{NullGMem2}
\end{equation} 
Note that despite the fact that the denomenator can vanish, the integrand in eqn. (\ref{NullGMem2}) is non-singular, because the denominator only vanishes when ${{\hat r}_y}={\hat r}$ and that is also where the numerator vanishes.  One can show that for the case where $F({{\hat r}_y})$ is any combination of $\ell=0$ and $\ell=1$ spherical harmonics, the integral in eqn. (\ref{NullGMem2}) vanishes.  Thus it is only the $\ell >1$ spherical harmonics that contribute to the memory.

Finally applying the results of eqn. (\ref{body2}) and eqn. (\ref{NullGMem2}) to eqn. (\ref{Gmem1}) we find that the total gravitational wave memory is 
\begin{eqnarray}
{M_{ij}} = {\frac 2 r} \; 
\biggl [  \; \int \; d {\Omega _y} \; {\frac {F({{\hat r}_y}) \, ( {P_{im}}{P_{jn}}  -  {\textstyle {\frac 1 2}} {P_{mn}}{P_{ij}}) {{\hat r}_y ^m}{{\hat r}_y ^n} } {1 - {\hat r}\cdot {{\hat r}_y}}} 
\nonumber
\\
+ \; {\sum _{\rm late}} {\frac {{m_{(\ell)}} {\gamma _{(\ell)}}( {P_{im}}{P_{jn}}  -  {\textstyle {\frac 1 2}} {P_{mn}}{P_{ij}}) {v_{(\ell)}^m}{v_{(\ell)}^n}} { 1 - {\hat r}\cdot {\vec v}_{(\ell)}}} \; 
\nonumber
\\
- \; {\sum _{\rm early}} {\frac {{m_{(\ell)}} {\gamma _{(\ell)}} ( {P_{im}}{P_{jn}}  -  {\textstyle {\frac 1 2}} {P_{mn}}{P_{ij}}) {v_{(\ell)}^m}{v_{(\ell)}^n}} { 1 - {\hat r}\cdot {\vec v}_{(\ell)}}} \biggr ] \; \; \; .
\label{Gmem2}
\end{eqnarray}
Here the first sum is over all late time isolated bodies with masses $m_{(\ell)}$ and velocities ${\vec v}_{(\ell)}$, while the second sum is the corresponding sum for the early time isolated bodies.

\section{Conclusion}
\label{Conclude}

We now contrast the approach of this paper to the approaches of \cite{thorne} and \cite{strominger}.  As noted {\it {e.g.}} in \cite{eannaandscott} the transverse-traceless part of the metric perturbation (which is the basis of the approach of \cite{thorne}) is one of a set of variables that is gauge invariant, but non-local.  In contrast, the perturbed Riemann tensor, which we use in this paper, is both gauge invariant and local.  Why would one go to the trouble of introducing and using a set of non-local gauge invariant variables when there is a simple and obvious set of local gauge invariant variables that one can use instead? Historically, I think the choice was made for ease of calculation: the metric has fewer indicies than the Riemann tensor.  Nonetheless, it is not widely understood that a choice has been made, and that there {\emph {is}} another (and conceptually simpler) way to treat gravitational radiation.  Therefore the calculational simplicity of the usual approach to gravitational radiation has been purchased at the cost of a certain conceptual confusion.

One can think of the metric perturbation $h_{\alpha \beta}$ as a potential for the perturbed Riemann tensor $R_{\alpha \beta \gamma \delta}$, analogous to the way that the electromagnetic vector potential $A_\alpha$ is a potential for the Farady tensor $F_{\alpha \beta}$.  Thus one way to express the difference between our approach and that of \cite{strominger} is that we emphasize field strengths ($F_{\alpha \beta}$ and $R_{\alpha \beta \gamma \delta})$, while \cite{strominger} emphasizes potentials and their gauge groups.  This emphasis on potentials and gauge groups is entirely appropriate for quantum field theory, and indeed one of the main purposes of \cite{strominger} is to point out relations between gravitational wave memory and certain concepts in quantum field theory.

This bridge building between general relativity and quantum field theory is all well and good.  However, it should be noted that quantum field theory is both more complicated and less well defined than (classical) general relativity.  (And quantum gravity is far more complicated and far less well defined than classical general relativity).  One should therefore not expect to find the best explanation of phenomena in the less abstruse theory by using the terms and methods of the more abstruse one.  Rather one should expect the reverse.  For example, \cite{strominger} notes a connection between gravitational wave memory and the infrared divergences of quantum field theory.  This connection is easily explained as follows: quantum field theory involves Fourier transforming everything, even when the Fourier transforms may not exist.  In particular, a smooth function $f(t)$ with a memory ({\it {i.e.}} with two different limits $f_+$ and $f_-$ as $t \to \pm \infty$), has a Fourier transform $g(k)$ for all nonzero $k$, and that Fourier transform diverges as $k \to 0$.  In this case memory provides a simple explanation for infrared divergences: {\emph {not}} the other way around.

In summary, this paper provides a conceptually simple explanation and perturbative derivation of electromagnetic and gravitational wave memory.  The treatment uses no non-local variables, no potentials of any kind, no gauge transformations and no gauge groups, because all these things are entirely unnecessary in this case.  Instead the treatment uses only the flat spacetime wave equation and the structure of the field equations, when expressed in terms of local gauge invariant variables.

\section*{Acknowledgments}

It is a pleasure to thank Lydia Bieri and Ratindranath Akhoury for helpful discussions.  This work was supported by NSF Grants PHY-1806219 and PHY-2102914. 

\appendix

\section{No gravitational kick}
\label{nokick}

In this section we show that the integral over both early and late null planes of ${Q^\mu}{n_\mu}$ with $Q^\mu$ given by eqn. (\ref{RiemannSource}) vanishes.  This will both validate expression (\ref{News2}) for the velocity tensor and show that the velocity tensor vanishes at late times.  Note that since a nonvanishing velocity tensor at late time would mean that there is a gravitatioinal kick, the main result of this section is that there is no gravitational kick.  Also note that it is sufficient to show that the integral over the null plane of terms of the form ${\partial _\alpha}{R_{\beta \gamma}}$ vanishes, since the integral of ${Q^\mu}{n_\mu}$ over the null plane consists of terms of that form.  Recall that at early times the stress-energy is that of isolated bodies, each travelling at constant velocity, whereas at late times, the stress energy consists of both such isolated bodies and outgoing radiation.  Consider the contribution of a single isolated body and adopt the rest frame of the body to perform the calculation.  Then in this rest frame there is no time dependence of $R_{\beta \gamma}$.  This means both that ${\partial _0}{R_{\beta \gamma}}=0$ and that the integral over the null plane is the same as the integral over space in the rest frame.  But then the integral over the null plane of ${\partial _i}{R_{\beta \gamma}}$ is equal to the integral over space of ${\partial _i}{R_{\beta \gamma}}$ which is easily seen to vanish simply by performing the integral over the $i$ coordinate first.  This validates eqn. (\ref{News2}) and shows that the contribution to the velocity tensor at late times from each isolated body vanishes.  

What remains is to show that the contribution of the radiation to the late time velocity tensor vanishes.  From eqn. (\ref{TNull}) it follows that for the radiation at late times 
\begin{equation}
{\frac 1 {4\pi}} \, {R^{\mu \nu}} = 2 {r_y ^{-2}} L ({t_r}-{r_y},{{\hat r}_y}) {\ell ^\mu} {\ell ^\nu}\; \; \; .
\label{Rnull}
\end{equation}
Since there is only a finite amount of radiated energy, the function $L$ vanishes at large positive and negative values of its first argument.  Furthermore, since the null plane ${\cal P}_{u_2}$ is given by ${t_r} -{r_y} = {u_2} - {{\hat r}_y}(1 - {\hat r}\cdot {{\hat r}_y})$, it follows that since $u_2$ is large, we must also use large $r_y$.  It then follows that for the integral over the null plane, we only need to compute ${\partial _\alpha}{R^{\mu \nu}}$ to order ${r_y ^{-2}}$.  However we have ${\partial _i}({r_y ^{-2}})= -2 {r_y ^{-3}} {{\hat r}_{yi}}$ and 
${\partial _i}{{\hat r}_{yj}} = {r_y ^{-1}}({\delta _{ij}} - {{\hat r}_{yi}}{{\hat r}_{yj}})$. It then follows from eqn. (\ref{Rnull}) that to order ${r_y ^{-2}}$ we have 
\begin{equation}
{\frac 1 {4\pi}} \, {\partial_\alpha} {R^{\mu \nu}} = - 2 {r_y ^{-2}} {\dot L} ({t_r}-{r_y},{{\hat r}_y}) {\ell _\alpha} {\ell ^\mu} {\ell ^\nu}\; \; \; .
\label{partialRnull}
\end{equation}
Here $\dot L$ denotes derivative of $L$ with respect to its first argument. 
It then follows that 
\begin{equation}
{\int _{{\cal P}_{u_2}}} \; {\frac 1 {4\pi}} \, {\partial_\alpha} {R^{\mu \nu}} = - 2 
\int \; d {\Omega _y} \; {\int _0 ^\infty} {r_y ^2} \; d {r_y} \; 
{r_y ^{-2}} {\dot L} ({u_2} - {{\hat r}_y}(1 - {\hat r}\cdot {{\hat r}_y}),{{\hat r}_y}) {\ell _\alpha} {\ell ^\mu} {\ell ^\nu}\; \; \; .
\label{IntpartialRnull}
\end{equation}
Changing variables to $ s \equiv {u_2} - {{\hat r}_y}(1 - {\hat r}\cdot {{\hat r}_y})$ we obtain
\begin{equation}
{\int _{{\cal P}_{u_2}}} \; {\frac 1 {4\pi}} \, {\partial_\alpha} {R^{\mu \nu}} = - 2 
\int \; d {\Omega _y} \; {\int _{-\infty} ^\infty}  \; ds \; 
{\frac { {\dot L} (s,{{\hat r}_y}) {\ell _\alpha} {\ell ^\mu} {\ell ^\nu}} {1 - {\hat r}\cdot {{\hat r}_y}}} \; \; \; .
\label{IntpartialRnull2}
\end{equation}
Then performing the integral over $s$ in eqn. (\ref{IntpartialRnull2}) we obtain zero, since the integral of $\dot L$ is $L$ and since $L$ vanishes at large positive and negative values of its argument.  Thus all contributions to the velocity tensor at large times vanish, which in turn implies that the velocity tensor at large times vanishes.  There is therefore no gravitational kick.

Note that the absence of a gravitational kick is a property of asymptotically flat spacetimes.  For example, as shown in \cite{gibbons} exact nonlinear plane waves provide a nonzero gravitational kick.

\end{document}